\documentclass[varenna]{cimento}

\usepackage{graphicx}
\usepackage{dcolumn}
\usepackage{bm}

\usepackage[utf8]{inputenc}
\usepackage[T1]{fontenc}
\usepackage{amsmath}
\usepackage{amssymb}
\usepackage{bm}
\usepackage{amssymb}
\newcommand{\be}{\begin{equation}}
\newcommand{\ee}{\end{equation}}
\newcommand{\bea}{\begin{eqnarray}}
\newcommand{\eea}{\end{eqnarray}}
\newcommand{\lan}{\langle}
\newcommand{\ran}{\rangle}

\title{Reaction Studies of Lepton Number Violation}

\author{H. Lenske} 
\institute{
University of Giessen, Giessen, Germany and NUMEN Collaboration, LNS Catania
}

\begin{document}



\maketitle              

\begin{abstract}
Nuclear isotensor spectroscopy as accessible in nuclear double charge exchange (DCE) reactions is indispensable for quantitative studies of lepton number violation as in double beta decay (DBD).
For such studies heavy ion double single charge exchange (DSCE) and direct Majorana double charge exchange (MDCE) reactions are discussed.  Isotensor two-body transition densities are investigated for the first time. Pion-potentials, mirroring neutrino potentials, and isotensor short range correlations are explored. Lepton DCE (LDCE) reactions on nuclei at accelerators are introduced as a promising new approach to investigate lepton number violation.
\end{abstract}

\section{Introduction}\label{sec:Intro}

Heavy ion double charge exchange (DCE) reactions proceeding by collisional interactions are the ideal tool to investigate the hitherto rarely studied and only rudimentarily understood isotensor response of nuclei. Double single charge exchange (DSCE) by acting twice with the nucleon-nucleon (NN) isovector T-matrix $\mathcal{T}_{NN}$ \cite{Lenske:2021jnr,Lenske:2024dsc} and single step Majorana DCE (MDCE) by acting in each nucleus twice with the pion-nucleon T-matrix $\mathcal{T}_{\pi N}$ \cite{Lenske:2024mdc} are the appropriate mechanisms for
spectroscopic studies. Experimentally, DCE reactions are being investigated as part of the NUMEN project \cite{Cappuzzello:2022ton} and independently at RCNP and RIKEN \cite{Takahisa:2017xry,Sakaue:2024qdc}, respectively. DSCE and MDCE probe nuclear isotensor modes give access to the nuclear configurations encountered in double beta decay (DBD).
Neutrinoless Majorana DBD (MDBD) is of special interest. At observation, MDBD will be a signature of  lepton number violation (LNV), indicating physics beyond the standard model (BSM), see e.g. \cite{GERDA:2023wbr}.
A hitherto neglected alternative, bridging the gap between the low-energy nuclear and high-energy collisional LNV experiments, are ($e^-,e^+$) LDCE reactions on nuclei ubder full laboratory control at accelerators. They are explored theoretically below and first estimates of cross sections are presented.

\begin{figure}
\begin{center}
\includegraphics[width = 5.5cm]{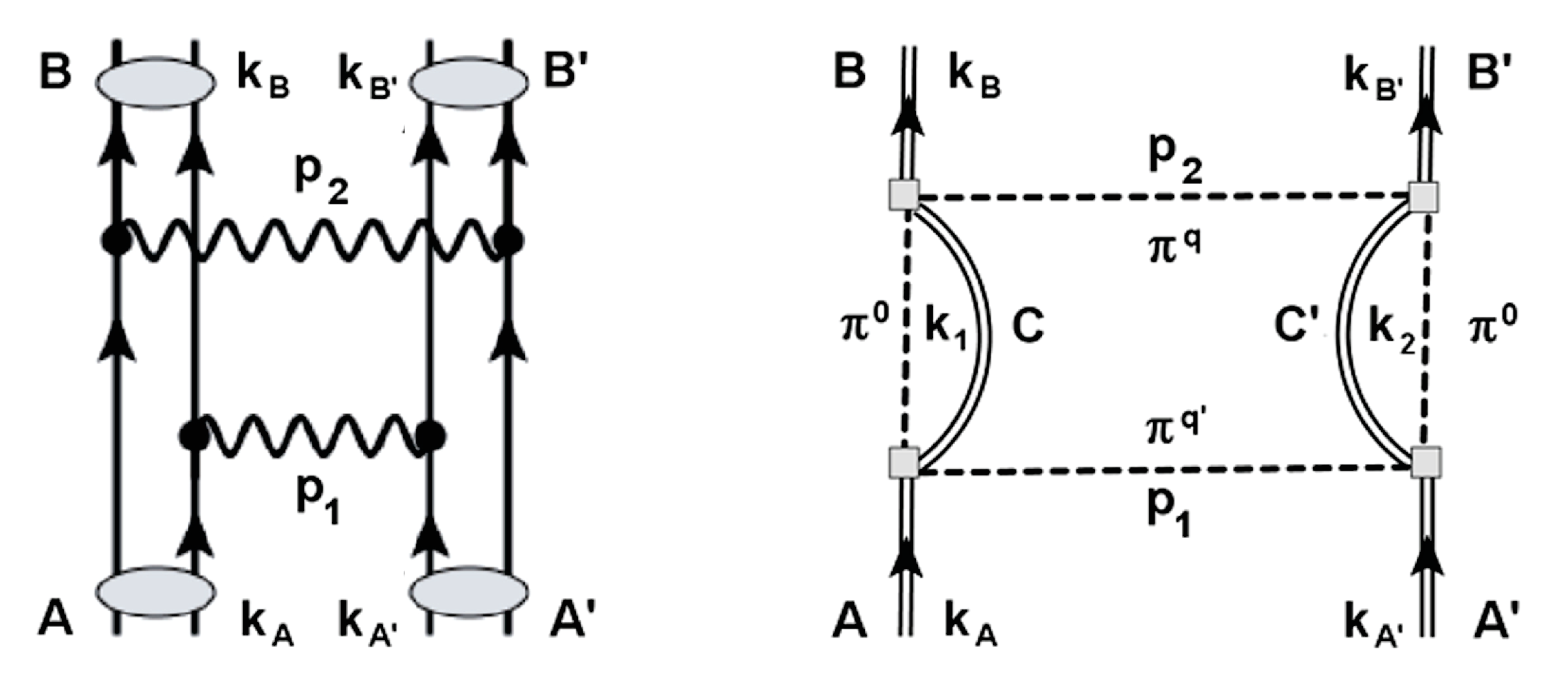}
\caption{Feynman diagrams of the DSCE (left) and the MDCE (right) reaction mechanisms contributing to a DCE reaction $A+A'\to B+ B'$ and the
respective four-momenta $k_{A,B}$ and $k_{A',B'}$. NN isovector T-matrices are indicated by filled circles (left), pion-nucleon T-matrices are depicted by grey boxes (right).}
\label{fig:DSCE_MDCE}
\end{center}
\end{figure}

\section{Nuclear Double Charge Exchange Reactions}\label{sec:DCE}

From the DSCE and MDCE Feynman diagrams, depicted in Fig.\ref{fig:DSCE_MDCE}, it is easily concluded that
the DSCE mechanism is a sequential process of second order in the NN isovector T-matrix $\mathcal{T}_{NN}$, the latter discussed e.g. in \cite{Love:1981gb,Franey:1985ye,Love:1987zx}. In contrast, MDCE reactions are first order reactions with respect to ion-ion interactions but intrinsically to each nucleus
they are of second order in the pion-nucleon isovector T-matrix $\mathcal{T}_{\pi N}$.
DSCE and MDCE theory and applications are well documented in \cite{Lenske:2021jnr,Lenske:2024dsc,Lenske:2024mdc,Bellone:2020lal}. Heavy ion DCE physics explores reactions between an entrance channel $\alpha=A+A'$ and exit channel $\beta=B+B'$ where the nuclear charges are changed by $\pm 2$ units but conserving the total charge. The reordering of charge  corresponds to rotations in isospin space. The different mechanisms are reflected in the formal structure of the MDCE, Eq.\eqref{eq:MMDCE}. and the DSCE, Eq.\eqref{eq:MDSCE}, amplitude:
\bea
&&\mathcal{M}^{(1)}_{\alpha\beta}(\mathbf{k}_\alpha,\mathbf{k}_{\beta})=
\lan \chi^{(-)}_\beta|\mathcal{W}_{AB}D_{\pi\pi}\mathcal{W}_{A'B'}|\chi^{(+)}_{\alpha} \ran \label{eq:MMDCE},\\
&&\mathcal{M}^{(2)}_{\alpha\beta}(\mathbf{k}_\alpha,\mathbf{k}_{\beta})=\lan \chi^{(-)}_\beta, BB'|\mathcal{T}_{NN}\mathcal{G}^{(+)}_{aA}(\omega_\alpha)\mathcal{T}_{NN}|AA',\chi^{(+)}_{\alpha} \ran  \label{eq:MDSCE}.
\eea
The MDCE amplitude is given by nuclear two-body isotensor transition form factors $W_{AB}$ and $W_{A'B'}$, respectively, and exchange of a pair of charged pions described by the propagator $D_{\pi\pi}$. Hence, each nucleus has its own set of intrinsic s-channel isotensor operators, generated and maintained by the t-channel exchange of a pair of pions in an $I=2$ isospin configuration. MDCE dynamics is determined by the isovector pion-nucleon T-matrix, dominated by the excitation of elastic resonances like $\Delta_{33}(1232)$ and higher lying P-wave and S-wave states \cite{Lenske:2024mdc} which affect the strengths at low energies and below threshold.

\begin{figure}
\begin{center}
\includegraphics[width = 8.5cm]{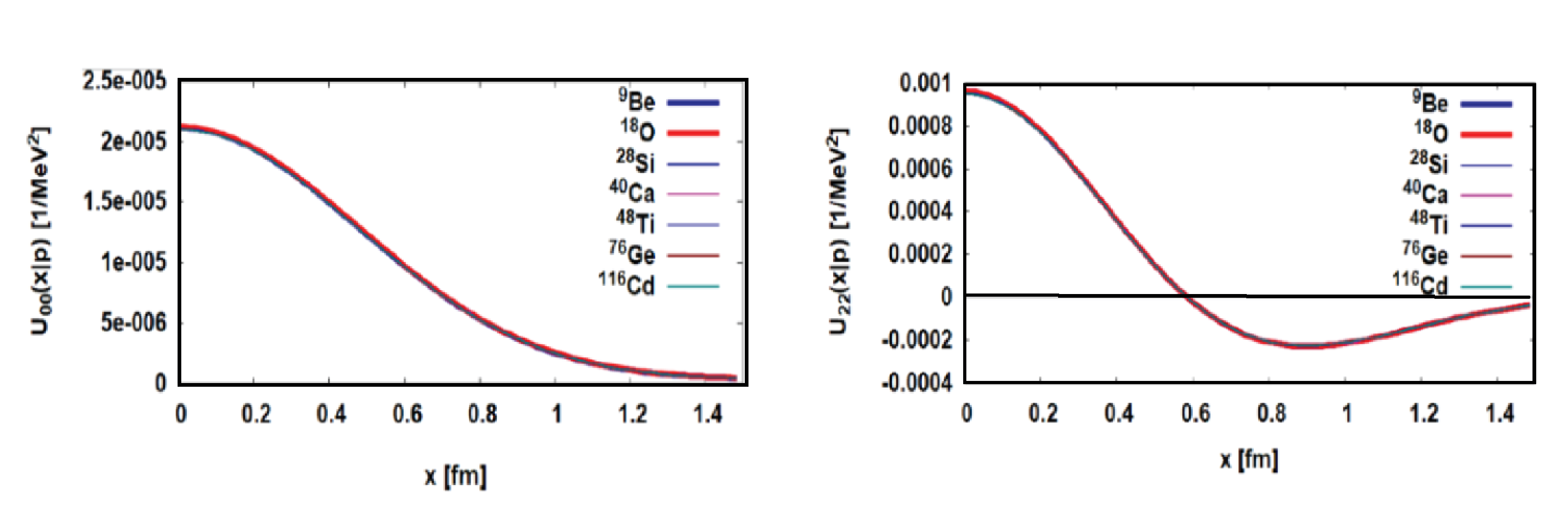}
\caption{The MDCE pion potentials $U_{00}$ and $U_{22}$ defining the strength of the double-Fermi and the double-Gamow-Teller components, evaluated in the collinear limits for $p_1=p_2=400$~MeV(c. Magnitude and shape are almost independent of the external momenta and the nuclear masses (from \cite{Garofalo:2024mth}).}
\label{fig:PionPot}
\end{center}
\end{figure}

Since the pion rest mass $m_\pi$ defines a natural separation scale, the MDCE amplitude can safely be evaluated in closure approximation leading to the  pion potential \cite{Lenske:2024mdc}
\be \label{eq:PionPot}
\mathcal{U}_\pi(\mathbf{x}|\mathbf{p}_{1,2}\bm{\sigma}_{1,2})=-\int \frac{d^3k}{(2\pi)^3}
\mathcal{T}_{\pi N}(\mathbf{p}_2,\mathbf{k}|\bm{\sigma}_2)\frac{e^{i\mathbf{k}\cdot \mathbf{x}}}{k^2+m^2_{\pi^0}}\mathcal{T}_{\pi N}(\mathbf{p}_1,\mathbf{k}|\bm{\sigma}_1).
\ee
The product of pion-nucleon T-matrices serves to construct the nuclear rank--2 isotensor $\mathcal{T}_{2\pm 2}=\left[\bm{\tau}_1\otimes \bm{\tau}_2\right]_{2\pm 2}$ and accordingly in the pion-sector.
$\mathcal{U}_\pi(\mathbf{x})$ consists in general of nine terms. In collinear kinematics $\mathbf{p}_1 || \mathbf{p}_2$, however, only six independent scalar form factors, $U_{ij}(x|p_1,p_2)\sim T_iT_j$, $i\leq j=0,1,2$ are found. These form factors define the strengths of a spin-scalar double Fermi term, a mixed Fermi-Gamow-Teller spin-vector component, and a double Gamow-Teller term which splits up into a two-body spin-spin operator and a rank-2 spin tensor operator. Typical results for $U_{ij}$  are shown
In Fig.\ref{fig:PionPot} $U_{00}$ and $U_{22}$ are shown as functions of the distance $x$ between the two participating nuclei. The correlation lengths are of the order of the pion wave length, $\sqrt{\lan x^2\ran}\simeq 1$~fm, thus being of much shorter range than the \emph{long-range} DBD correlations from the neutrino potential.

The DSCE kernel
$\mathcal{K}_{\alpha\beta}=\mathcal{T}_{NN}\mathcal{G}^{(+)}_{aA}(\omega_\alpha)\mathcal{T}_{NN}$
acts as effective isotensor interactions which in this case is achieved by two separate t-channel isovector interactions. In \cite{Lenske:2021jnr,Lenske:2024dsc} that operator was investigated in detail by deriving spin-angular momentum multipole decompositions. Reaction-theoretically the DSCE process is described in t-channel formulation as can be deduced from the ladder-type DSCE diagram in Fig.\ref{fig:DSCE_MDCE}.  For spectroscopic studies, however, the s-channel representation is the appropriate formulation. In \cite{Lenske:2024dsc} the kernel is transformed into a superposition of two-body isotensor operators acting intrinsically in $A$ and $A'$.

\begin{figure}
\begin{center}
\includegraphics[width = 6.5cm]{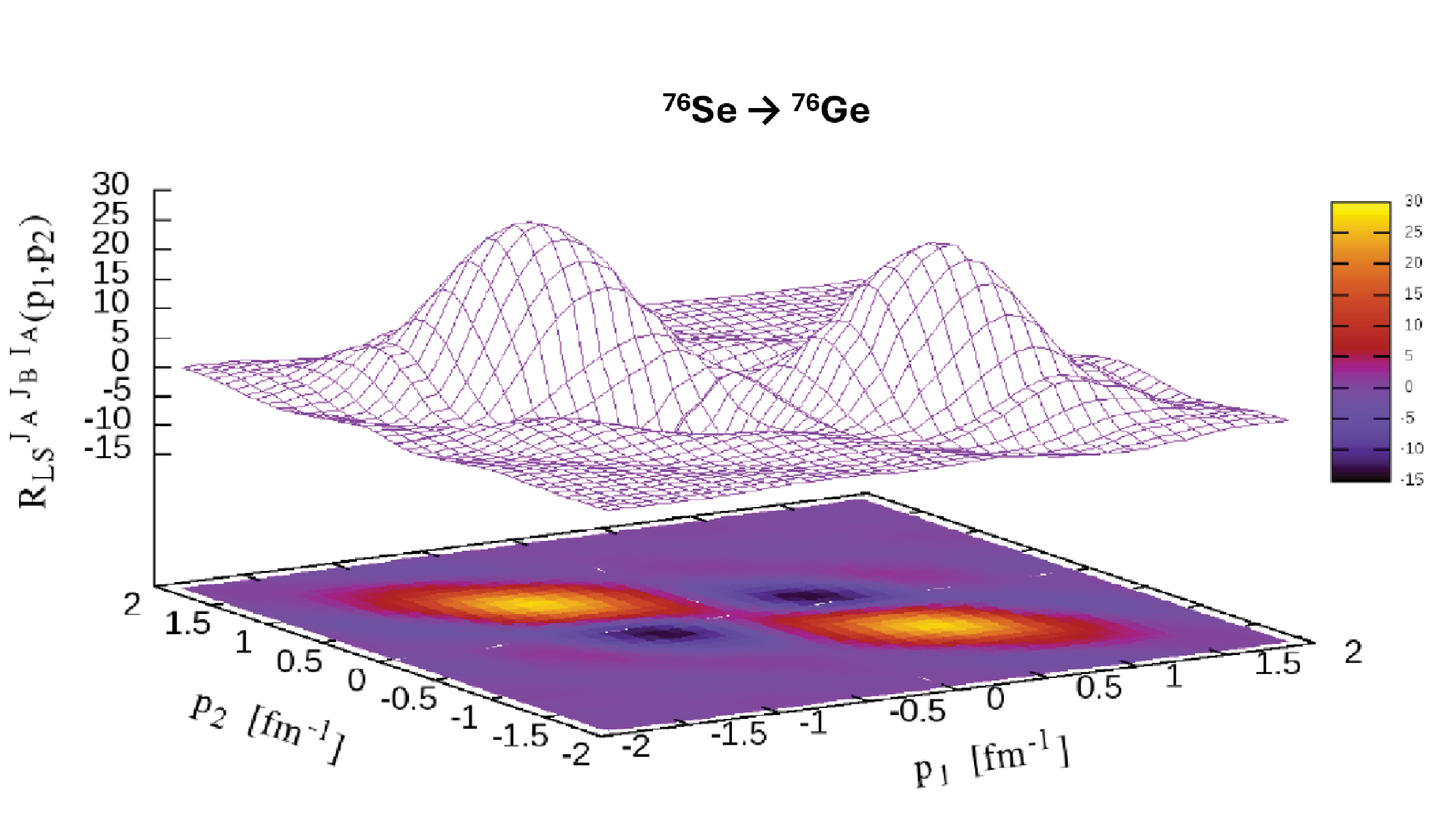}
\caption{Momentum space isotensor two-body transition density of the double-Fermi spin-scalar component for the DCE transition $^{76}$Se$\to{}^{76}$Ge as function of the momenta $\mathbf{p}_{1,2}$ transferred in the 1$^{st}$  and 2$^{nd}$ SCE step, projected to the total momentum transfer $\mathbf{q}$. Note that the momenta extend over $\pm 400$~MeV/c (from \cite{Bellone:2025ada}).}
\label{fig:TBTD}
\end{center}
\end{figure}

In s-channel representation, advantage is taken of the unique feature of collisional DCE reactions to probing directly isotensor component of the nuclear two particle-two hole ($N'^2N^{-2}$) response. In \cite{Bellone:2025ada} a first-time-ever study of a $n^2p^{-2}$ two-body transition density distributions (TBTD) was exercised for the DSCE reaction $^{76}Se({}^{18}O,{}^{18}Ne){}^{76}Ge$, hence addressing the inverse of the MDBD-candidate $^{76}Ge_{g.s.}\to{}^{76}Se_{g.s.}$.
In Fig.\ref{fig:TBTD} the TBTD for the double-Fermi component $S_1=S_2=0$ for the $A=76$ transition is displayed as function of the projections of the momenta $\mathbf{p}_{1,2}$, along the total momentum transfer $\mathbf{q}$ of the reaction. The TBTD for the double-Gamow-Teller spin-tensor part ($S_1=S_2=1$), coupled to $S=2$, and the mixed Fermi-Gamow-Teller TBTD for ($S_1=0,S_2=1$) are found in Ref.\cite{Bellone:2025ada}.

\section{Lepton Double Charge Exchange Reactions}

\begin{figure}
\begin{center}
\includegraphics[width = 7.5cm]{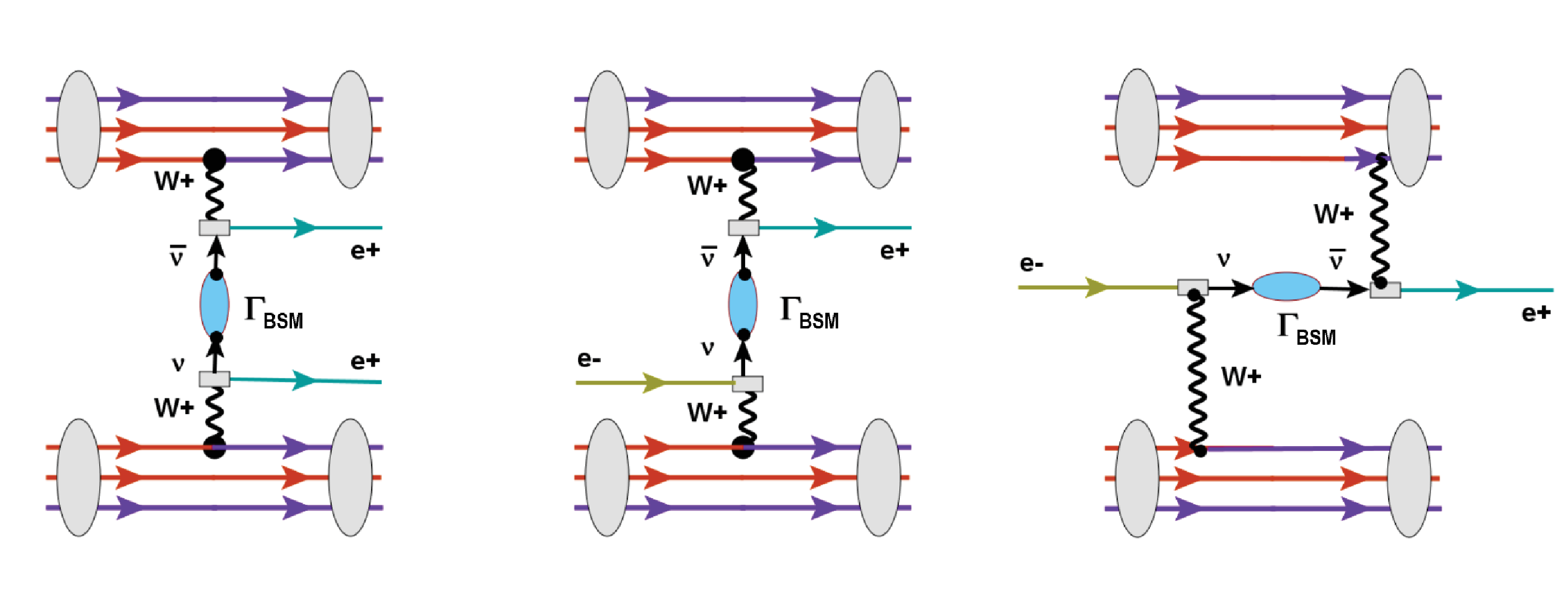}
\caption{The Feynman diagrams, with LNV-vertices $\Gamma_{BSM}$, for the t-channel capture (center) and the s-channel scattering (right) LDCE processes on a proton pair are compared to the MDBD decay diagram (left). Up- and down-quarks are denoted by red and blue lines, respectively.  }
\label{fig:LDCE}
\end{center}
\end{figure}

A tempting alternative to MDBD and collisional $\ell^2j^2$ events are LNV accelerator experiment, although never considered before. ($e^-,e^+$) experiments at nuclear targets can be regarded as available prototypes of such studies. In Fig.\ref{fig:LDCE} Feynman diagrams of the elementary processes are displayed for MDBD and the t-channel and the s-channel reaction counterparts. Surprisingly, nothing is known about such reactions, neither theoretically nor experimentally. The theory of the s-channel LDCE mechanism is based on the Left-Right Symmetric Model (LRSM). Within acceptable approximations
the LRSM cross sections are reducible to a form of second-order perturbation theory which allows  using (half off-shell) charged current (CC) ($e^-,\nu$) and ($\bar{\nu},e^+$) amplitudes are input. Since CC amplitudes over large incident energy ranges up to the deep-inelastic region are needed, a semi-phenomenological approach is the method of choice at this stage. The s-channel graph shown in Fig.\ref{fig:LDCE} is especially interesting because the lepton channel is populated according to the incident energy, thus probing directly the energy dependence of the LNV vertex together with the propagation of the intermediate system. The competing t-channel process probes primarily
the three-momentum structure of $A(e^\pm,\nu)C$ vertices which will not profit of high beam energies. In left (L) and right (R) helicity representation the LDCE amplitude is
\bea\label{eq:LDCE}
&&M^{(e^-e^+)}_{\alpha\beta}(\mathbf{k}_\alpha, \mathbf{k}_\beta|w_\alpha)\approx \sum_{\lambda,\lambda'=L,R}
\sum_{C}\\
&&\int^{T_\alpha}_0 dE \rho_C(E)\int\frac{d^3k}{(2\pi)^3}
M^{(e^+\bar{\nu};\lambda)}_{\beta\gamma}(\mathbf{k}_\beta,\mathbf{k})
\mathcal{S}^{(\lambda\lambda')}_{C}(\mathbf{k}|w_\alpha)
\widetilde{M}^{(\nu e^-;\lambda')}_{\gamma\alpha}(\mathbf{k},\mathbf{k}_\alpha),\nonumber
\eea
where $\lambda,\lambda'\in \{L,R\}$ and $\mathcal{S}^{(\lambda\lambda')}_{C}$ is the matrix of helicity-diagonal and helicity-mixing effective flavor propagators, obtained from intermediate propagators by averaging over the massive neutrinos including the
flavor-mass transformation matrices $U^{(\lambda)}_{ei}$. If there are only three light massive neutrinos, the PMNS matrices will be recovered.

\begin{figure}
\begin{center}
\includegraphics[width = 7cm]{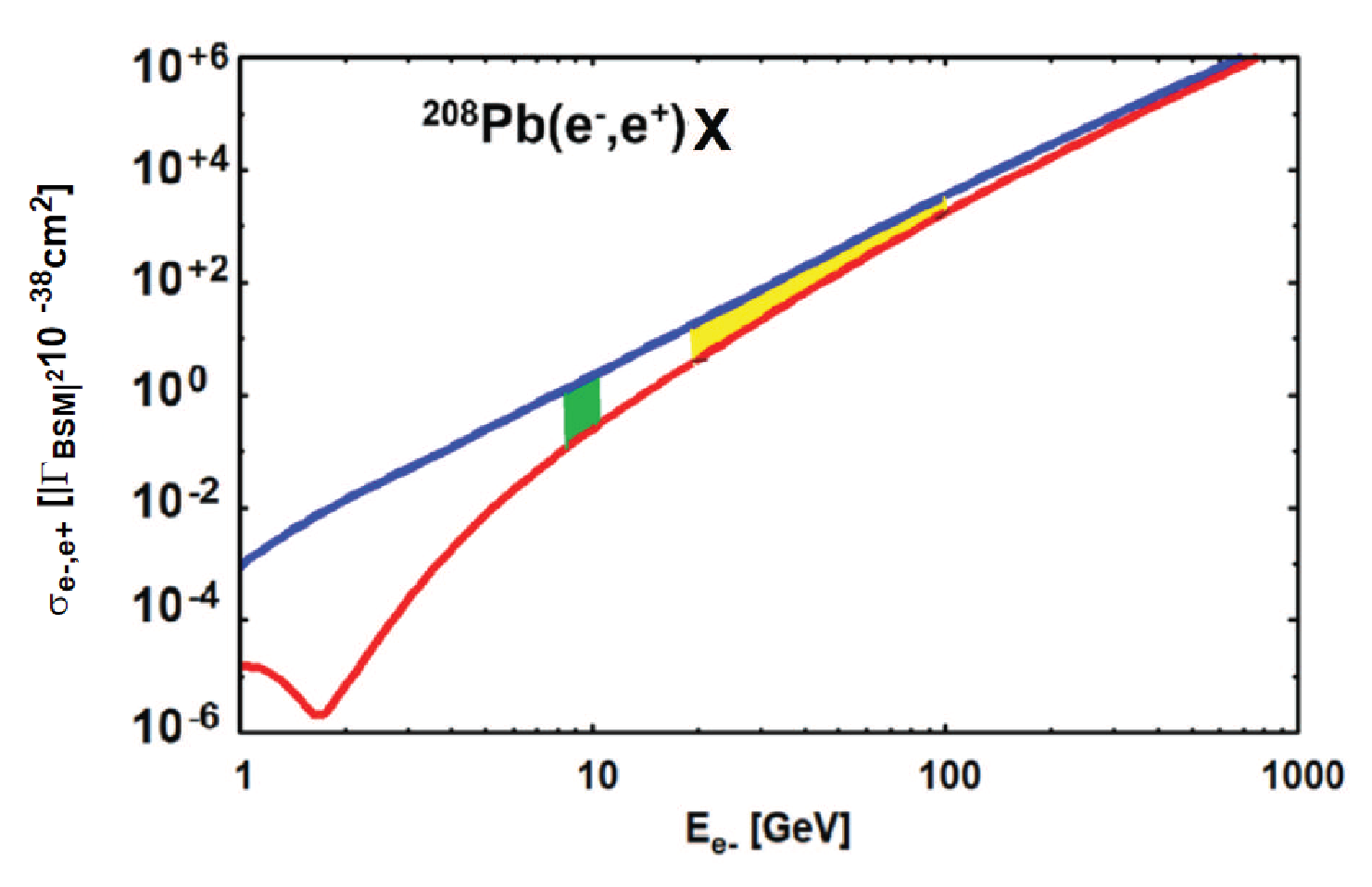}
\caption{LDCE total cross sections for $e^- + {}^{208}$Pb$\to e^+ + X$ where $X$ is constrained only by the total charge $Z_X=Z-2=80$ and fixed baryon number $A=208$. Cross sections are given in units of the unknown dimensionless LNV parameter $\Gamma_{BSM}$. Cross section in the energy range available at Jefferson Laboratory (green) and at the planned Electron-Ion-Collider EIC (yellow) are marked. The enclosed area indicates uncertainties on relative phases only.}
\label{fig:LDCEXS}
\end{center}
\end{figure}

Existing CC cross section studies are performed typically only up to the resonance region, $T_{lab}\lesssim 2$~GeV which was found too restrictive. In order to obtain dependable results on ($e^-,e^+$) cross sections over large ranges of incident energies, a phenomenological approach was chosen by extracting CC matrix elements from the compilations in \cite{Formaggio:2012cpf}.
Assuming universality, those amplitudes were used to evaluate numerically Eq.\eqref{eq:LDCE}. The derived total cross sections are increasing in energy $\sim E^3$ and grow rapidly with target mass. Thus, the most favorable case is to use heavy target nuclei and electron beams of the highest available energy. In Fig.\ref{fig:LDCEXS} total cross section for ($e^-,e^+$) reactions on an $^{208}$Pb target are shown. At 10~GeV beam energy one finds $\sigma_{e^-e^+}\sim |\Gamma_{BSM}|^2 10^{-38}cm^2$ in units of (dimensionless) LNV parameters, estimated theoretically as $\Gamma_{BSM} \lesssim 10^{-7}$.

\section{Summary}\label{sec:Sum}

The theory of heavy ion DCE reactions was discussed under the aspect to obtain a new access to nuclear isotensor spectroscopy. The sequential DSCE and the direct MDCE mechanisms provide the appropriate dynamical conditions. As a first spectroscopic application, the momentum structure of isotensor two-body transition densities was explored theoretically. MDCE dynamics is given by virtual pion-nucleon DCE scattering, leading to pion potentials which are playing the role of the neutrino potentials known from MDBD. As new approach to LNV studies high energy ($e^-,e^+$) reactions on nuclear targets were proposed. Total cross section for a lead-target were estimated in a phenomenological model.

\paragraph{\textbf{Acknowledgment:}} Support by DFG, grant Le439/17 and INFN--LNS Catania and collaboration with the NUMEN theory and experimental groups are gratefully acknowledged.
%

\end{document}